\documentclass[twocolumn]{aastex631}

\newcommand{\neiii}{[Ne\textsc{iii}]$\lambda$3869}
\newcommand{\oii}{[O\textsc{ii}]$\lambda\lambda$3727,3729}
\newcommand{\oiii}{[O\textsc{iii}]$\lambda\lambda$4959,5007}
\newcommand{\neon}{[Ne\textsc{iii}]}
\newcommand{\sm}{log(M_{\star}/M_{\odot})}

\begin{document}

\title{The Neon Gap: Probing Ionization with Dwarf Galaxies at z$\sim$1}

\correspondingauthor{John Pharo}
\email{jp7tp@missouri.edu}

\author{John Pharo}
\affiliation{Department of Physics and Astronomy, University of Missouri, Columbia, MO 65211, USA}

\author{Yicheng Guo}
\affiliation{Department of Physics and Astronomy, University of Missouri, Columbia, MO 65211, USA}

\author{David C. Koo}
\affiliation{UCO/Lick Observatory, Department of Astronomy and Astrophysics, University of California, Santa Cruz, CA, USA}
\affiliation{University of California, Santa Cruz, CA, USA}

\author{John C. Forbes}
\affiliation{Center for Computational Astrophysics at the Flatiron Institute, New York, NY, USA}

\author{Puragra Guhathakurta}
\affiliation{UCO/Lick Observatory, Department of Astronomy and Astrophysics, University of California, Santa Cruz, CA, USA}
\affiliation{University of California, Santa Cruz, CA, USA}



\begin{abstract}

We present measurements of \neiii\ emission in $z\sim1$ low-mass galaxies taken from the Keck/DEIMOS spectroscopic surveys HALO7D and DEEPWinds. We identify 167 individual galaxies with significant \neon\ emission lines, including 112 ``dwarf'' galaxies with $\sm <9.5$, with $0.3<z<1.4$. We also measure \neon\ emission from composite spectra derived from all \oii\ line emitters in this range. This provides a unique sample of \neon-emitters in the gap between well-studied emitters at $z=0$ and $2<z<3$. To study evolution in ionization conditions in the ISM over this time, we analyze the log(\neiii/\oii) ratio (Ne3O2) as a function of the stellar mass and of the log(\oiii/\oii) ratio (O32). We find that the typical star-forming dwarf galaxy at this redshift, as measured from the composite spectra, shares the Ne3O2-$M_{\star}$ relation with local galaxies, but have higher O32 at given Ne3O2. This finding implies that the ionization and metallicity characteristics of the $z\sim1$ dwarf population do not evolve substantially from $z\sim1$ to $z=0$, suggesting that the known evolution in those parameter from $z\sim2$ has largely taken place by $z\sim1$. Individual \neon-detected galaxies have emission characteristics situated between local and $z\sim2$ galaxies, with elevated Ne3O2 and O32 emission potentially explained by variations in stellar and nebular metallicity. We also compare our dwarf sample to similarly low-mass $z>7$ galaxies identified in JWST Early Release Observations, finding four HALO7D dwarfs with similar size, metallicity, and star formation properties. 

\end{abstract}

\keywords{Galaxy evolution - emission line galaxies - interstellar medium}


\section{Introduction} \label{sec:intro}

The conditions of the gaseous interstellar medium (ISM) hosting star formation in galaxies may yield key insights into relationships between the local stellar populations and the physical processes that influence the level of star formation in galaxies. It has been well-established that the level of cosmic star formation has evolved over time, reaching a peak around $z\sim2$, after which cosmic star formation declines \citep{madau14}. Comparison of galaxies in this epoch of peak star formation with local galaxies has suggested a corresponding evolution in the ionization conditions of the gas-phase medium for galaxies at fixed stellar mass \citep{steidel2014, sanders2016, strom2017}. 

Empirical observations of emission line ratios of higher redshift galaxy populations show significant offsets compared with lower redshift \citep{steidel2014, sanders2016}. Proposed explanations for this offset include harder ionizing sources in higher redshift galaxies, requiring changes in the constituent stellar populations and/or nebular gas content of star forming galaxies across cosmic time \citep{strom2017, shapley2019} , or evolution  in the relation of stellar mass, gas-phase metallicity, and star formation \citep{sanders2018}. However, high-redshift studies typically have limited observations of low-mass (or ``dwarf'') galaxies, despite their numerical importance to the overall star-forming galaxy population \citep{muzzin13}, so the nature and timescale of this evolution can still be further constrained by additional observations. As JWST enables astronomers to push the study of ionizing conditions in low-mass galaxies to higher redshift, understanding the nature of the $z\sim2$ to $z=0$ evolution in the dwarf population becomes critical for characterizing this ISM history.

Restframe-optical emission lines are commonly used to study properties of the ISM in galaxies, including gas-phase metallicity (Z), electron temperature ($T_e$), element ionization state, instantaneous star formation rate (SFR), and dust content \citep[e.g.][]{kennicutt98,kewley01,kewley19,cardelli1989}. The most commonly used lines for this purpose typically include the intrinsically strongest Balmer lines (primarily H$\alpha$ and H$\beta$), as well as strong transitions of ionized metal lines (\oii; \oiii; [N\textsc{ii}]$\lambda$6584; [S\textsc{ii}]$\lambda\lambda$6717,6713), which have well-tested calibrations to SFR, Z, and other galaxy properties. However, these lines span a wavelength range of several thousand Angstroms, pushing redder lines into the near-infrared as redshift increases, as well as leaving line ratio measurements susceptible to uncertainties in dust extinction. 

The \neiii\ emission line may provide a useful alternative to strong lines at redder wavelengths. The \neiii/\oii\ ratio (Ne3O2) provides a monotonic metallicity diagnostic \citep{nagao2006}, as compared to double-branched diagnostics such as R23, and the lines' closee wavelength proximity renders the ratio insensitive to dust, unlike the similarly monotonic \oiii/\oii\ diagnostic. Ne3O2 is also a sensitive diagnostic of ionization parameter \citep{levesque2014}, and the short wavelengths of its constituent emission lines make it a valuable empirical measure for high-redshift studies. \citet{zeimann2015} first noted the possibility of enhanced Ne3O2 in Hubble Space Telescope grism spectroscopy of $z\sim2$ low-mass galaxies, and \citet{strom2017} and \citet{jeong2020} measured the ratio for more massive $z\sim2$ galaxies. However, Ne3O2 is not well studied at intermediate redshifts between the local universe and the $z\sim2$ peak of cosmic SFR. The line is relatively faint, requiring deep observations, and may be blended with nearby Helium and Hydrogen lines at 3888\AA\ in low-resolution spectra.

With deep Keck/DEIMOS spectra from the HALO7D, DEEPWinds, and other surveys, we are able to measure \neiii\ for 167 individual galaxies with $0.3<z<1.4$, including 112 galaxies below $\sm = 9.5$, which we define to be dwarf galaxies. In this paper, we use this sample to explore ionization characteristics and their evolution in the low-mass population in this transitional period.

This paper is organized as follows. In \S \ref{sec:data}, we describe the spectral data used in our analysis, including HALO7D and other studies used for comparison. We also describe measurement of the \neiii\ sample and our method for deriving composite spectra. \S \ref{sec:results} presents the empirical results of our measurements, while \S \ref{sec:disc} discusses the implications of the Ne3O2 distributions in the dwarf galaxy population and for the evolution of ionizing conditions in the ISM.

In this paper, we adopt a flat $\Lambda$CDM cosmology with $ \Omega_m = 0.3$, $\Omega_{\Lambda} = 0.7$, and the Hubble constant $H_0 = 70$ km s$^{-1}$ Mpc$^{-1}$. We use the AB magnitude scale \citep{oke83}.

\section{Data and Sample Description} \label{sec:data}

\subsection{HALO7D and Related Surveys}

The data used in this paper are comprised of deep optical spectra of $\sim$2400 galaxies observed with Keck/DEIMOS. Most spectra were taken by the HALO7D program \citep[PI: Guhathakurta;][]{cunningham2019a,cunningham2019b}, a program primarily designed to observe faint Milky Way halo stars in the COSMOS, EGS, and GOODS-North CANDELS fields \citep{grogin11, koekemoer11}. Unused space in the DEIMOS slit masks was filled out with galaxies, including a low-mass galaxy sample at $0 < z < 1.0$ in addition to high-mass galaxies targeted for studies of strong winds in star-forming galaxies \citep{wang2022}, AGN \citep{yesuf17}, and stellar populations in quiescent galaxies \citep{tacchella2022}. Additional programs expanded the sample to include GOODS-South. Dwarf galaxy targets were generally selected to have $0.4 < z < 0.9$, $7.0< \sm < 9.5$, and F606W mag $\leq26$. The total observations comprise a sample of 2444 target galaxies, including 1255 low-mass galaxies across four CANDELS fields, as well as 1189 more massive galaxies.

All spectra used in this project were obtained by the DEep Imaging Multi-Object Spectrograph (DEIMOS) instrument at the Keck Observatory \citep{faber03}. The Keck/DEIMOS spectrograph has an overall effective wavelength coverage of roughly $4100 < \lambda < 11000$ \AA. For the HALO7D observations, DEIMOS was configured with the 600 line mm$^{-1}$ grating centered at 7200\AA\, giving a wavelength dispersion of 0.65 \AA/pix, resolution $R\approx2100$, and a usable wavelength range limited to $5000 < \lambda < 9500$ \AA\ \citep{cunningham2019a}.

The observations were reduced using the automated DEEP2/DEIMOS \textit{spec2d} pipeline developed by the DEEP2 team \citep{cooperspec2d2012, newman13}, described fully in \citet{yesuf17} and \citet{cunningham2019a}. This yielded extracted 1D spectra for each exposure, and produced images of the reduced 2D spectra and extraction windows for the purposes of visual inspection of the data. The 2D spectra images were inspected for excessive contamination or other issues, and those exposures that passed visual inspection were coadded into a single 1D spectrum for each galaxy. The coadded spectra were then flux scaled to best-fit photometric SEDs. For more details on the coaddition and flux calibration of the spectra, see \citet{pharo2022}.

\begin{figure*}
    \centering
    \includegraphics[width=\textwidth]{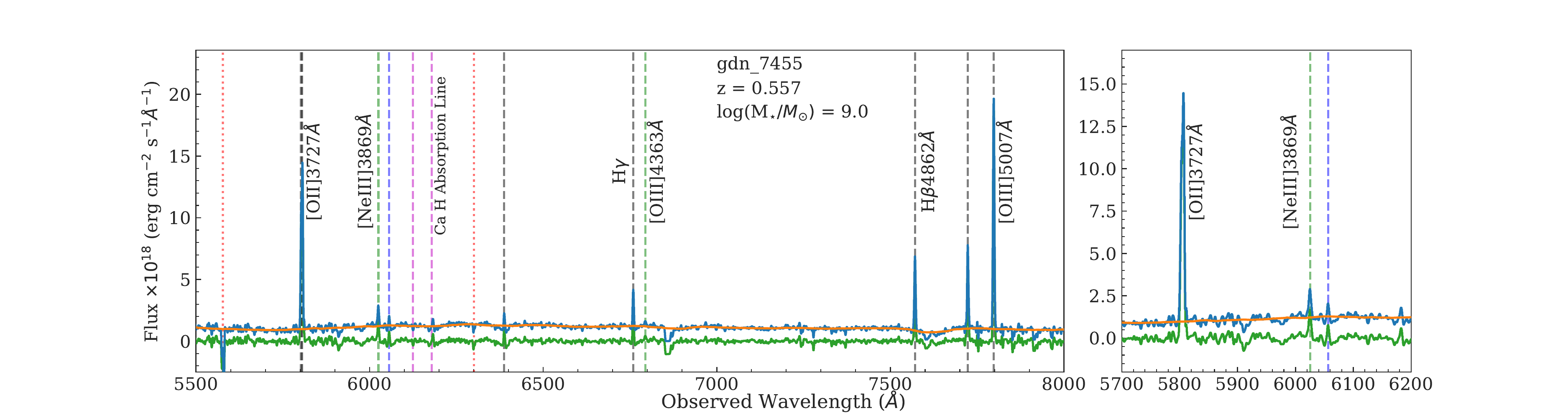}
    \caption{\textit{Left:} Example spectrum of a HALO7D dwarf emission line galaxy. The blue solid line gives the observed spectrum. The orange solid line shows the continuum estimate, and the green solid line the continuum-subtracted spectrum. Prominent emission lines are labelled and indicated by vertical dashed lines. Black dashed lines indicate strong and/or Balmer series emission lines. Green dashed lines indicate typically fainter ionized metal lines, and the blue dashed line shows the location of a blend of the faint HeI$\lambda$3889 and H$\zeta$ emission lines. The two magenta dashed lines show the locations of the Ca H and K stellar absorption lines, though these features are not prominent in this spectrum, which has little stellar continuum due to its low stellar mass as well as a likely young stellar population. \textit{Right:} A zoom-in on the spectrum showing the \oii\ doublet, as well as the \neiii\ emission line and HeI-H$\zeta$ blend.}
    \label{fig:exspec}
\end{figure*}

\begin{figure}
    \centering
    \includegraphics[width=0.5\textwidth]{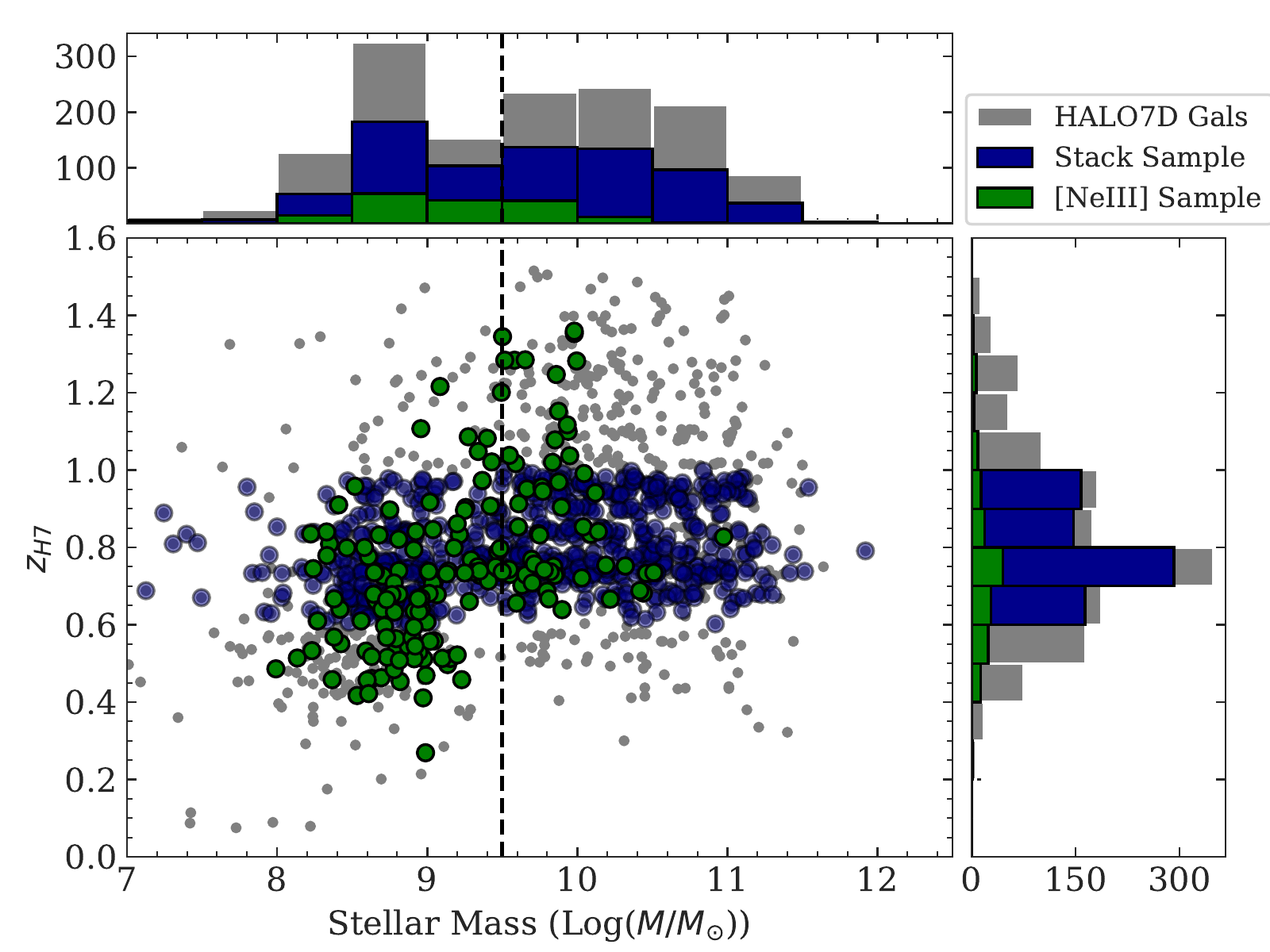}
    \caption{The stellar mass versus redshift distribution of the HALO7D galaxy sample. The vertical dashed line separates the dwarf and massive galaxy populations, defined at $\sm = 9.5$. The gray points and histograms indicate the overall sample of galaxies for which a good redshift fit was obtained from spectral features, as described in \citet{pharo2022}. The green points and histograms denote the subsample of galaxies with a significant (S/N$>3$) detections of the [Ne\textsc{iii}]3869 and [O\textsc{ii}]3727,3729 emission lines. This yields a sample of 167 galaxies (out of 1432 in the HALO7D catalog), including 112 dwarf galaxies (out of 646), with a redshift range of $0.3 < z < 1.4$. The dark blue points and histograms indicate the emission line galaxy sample used for composite galaxy stacking, described in \S2.4.}
    \label{fig:mz}
\end{figure}

\subsection{The \neiii\ Sample}

To obtain redshift measurements from the coadded 1D galaxy spectra, we developed a routine to fit strong emission lines in a $0 < z < 2$ redshift window, selected to encapsulate the region where strong line emitters were likely to be found. This procedure is described in detail in \citet{pharo2022}. In Figure \ref{fig:exspec}, we show an example HALO7D spectrum for a dwarf emission line galaxy, demonstrating many of these features. Emission lines that are both faint and rare, such as the \neiii\ emission line, require more careful attention to avoid false detections, and so are measured after redshift fitting is complete.

With the complete redshift catalog, we then visually inspected each spectrum for evidence of significant faint line emission, as well as fitting a list of faint lines. Visually-flagged galaxies where \neiii\ was measured with S/N$>3$ were added to a candidate list. See Figure \ref{fig:mz} for a visual of the redshift and stellar mass distributions of this subsample relative to the overall emission line galaxy catalog. The 167 individual \neiii\ detections represent only 11\% of the overall HALO7D emission line catalog (see \citet{pharo2022} for details). The \neiii-detected galaxies have a median cumulative observation time of $\sim8$ hours, representing observations longer by a factor of 8 than in the previous DEEP2 survey.

\subsection{Comparison Samples}

To investigate changes in \neon\ emission with redshift and mass, we make use of several existing studies that have measured similar emission lines. At $z\sim0$, we make use of composite spectra of SDSS emission line galaxies measured in \citet{am2013}, hereafter AM13. To include more low-mass galaxies at $z\sim0$, we add 38 galaxies from the Local Volume Legacy survey \citep[][, hereafter LVL]{berg2012}, 34 of which have \neiii\ detections. We create makeshift composite spectra for this sample by taking the median Ne3O2 ratios in bins of stellar mass, treating the nondetections as having ratios of 0. At higher redshift, we compare to three studies of star-forming galaxies: from the MOSDEF survey \citep{kriek2015}, where \citet[][hereafter J20]{jeong2020} measure Ne3O2 for 61 $z\sim2.3$ galaxies; KBSS-MOSFIRE \citep{rudie2012}, where \citet[][hereafter S17]{strom2017} measured emission line ratios for 69 $2 < z < 3$ galaxies; and composite measurements of low-mass galaxies from HST spectra in \citet{zeimann2015}. We also compare to recent JWST observations of three $z>7$ galaxies, adopting the stellar mass and emission line measurements done in \citet{schaerer2022}. We note that these measurements come from initial flux calibrations and SED fits, and so may change with further analysis.

\subsection{Composite Spectra}

The analysis of emission line measurements in individual galaxies is necessarily limited to those galaxies whose observations have the signal necessary to detect the emission lines. This introduces possible selection biases, in particular for samples of intrinsically faint emission lines. By combining groups of individual spectra into composite spectra, we may obtain average measurements for subsamples of galaxies that include galaxies without individual detections of a given emission line. This may then provide a more representative measure of the emission characteristics of that galaxy subsample.

For this stacking procedure, we limit our sample to dwarf galaxies with a significant [O\textsc{ii}]3727,3729\AA\AA\ detection in \citet{pharo2022}, so as to avoid including quiescent massive galaxies from the HALO7D sample \citep{tacchella2020}. This should not bias the sample, since [O\textsc{ii}]3727,3729\AA\AA\ is typically one of the brightest emission lines detected, and it is close in wavelength to [Ne\textsc{iii}]3869 so the redshift range will not be artificially restricted. 

After the sample is selected, we sort the galaxies by redshift and mass into bins containing comparable numbers of galaxies. Since the mass selection of galaxies in HALO7D is not uniform across redshift (see Figure \ref{fig:mz}), we restrict our redshift range to $0.6 < z < 1.0$, where the full range of masses has been sampled. The distribution of the stacked sample in Figure \ref{fig:mz} shows that the stacking sample is composed of the vast majority of HALO7D emission line galaxies in the given redshift range. \citet{pharo2022} investigated the star formation and color-magnitude properties of HALO7D emission line galaxies relative to CANDELS photometric catalogs, finding them comparable to CANDELS on the star-forming main sequence at similar redshift and with similar color properties. This indicates our stacking sample ought to be representative of the general galaxy population. In order to produce bins with enough constituent galaxies to yield a meaningful average, we use six mass bins. The details on the bin sizes and their constituent galaxies are described in the left and middle panels of Table \ref{tab:stack}. 

For each bin, we then combine the individual spectra with the following procedure. First, the continuum is estimated and subtracted using the median-filter method described in \S3.2 of \citet{pharo2022}, and the residual fluxes are normalized to the \oii\ line flux. We choose to initially remove the continuum from all galaxies in order to avoid difficulties with particularly low-mass galaxies, where the continuum is often not well-detected. Normalization to the \oii\ flux eliminates any issue of relative flux dimming from the slightly different redshifts among galaxies in the same bin, and since we are primarily concerned with emission line ratios rather than their fluxes, we may operate with this normalization. Since we are primarily concerned with the Ne3O2 ratio, we do not consider dust extinction, the effects of which should be minimal.

Next, the normalized spectrum is rebinned onto a uniform grid of wavelengths. Once each spectrum in the redshift-mass bin has been normalized and rebinned, they are stacked together by taking the median flux at each wavelength. Normalized emission line fluxes may then be measured from each composite spectrum, along with the median stellar absorption as a fraction of line emission. For lines with possibly significant stellar absorption, we measure the flux and absorption by simultaneously fitting emission and absorption profiles. To obtain errors for the emission line measurements, we use a Monte Carlo bootstrap method wherein the constituent galaxies of the bin are resampled with replacement 100 times, and the new samples stacked and emission lines measured 100 times. Each instance, we perturb the flux at each wavelength by sampling a Gaussian flux distribution centered on the flux measurement with a width set by the error spectrum. The errors for each emission line are then estimated from the standard deviation in the resulting distribution of measurements. We find typical Ne3O2 errors of $\sim$0.1 dex, comparable to those measured in AM13 and J20.

\begin{deluxetable*}{cc|cccc|cccc}
    \label{tab:stack}
    \centering
    \tablecaption{Composite Spectra Bins for Ne3O2 and O32 Stacks}
    \tablecolumns{5}
    \tablehead{
    \colhead{$z_{min}$} &
    \colhead{$z_{max}$} \vline& 
    \colhead{Median $z$} &
    \colhead{Median Mass} &
    \colhead{N} &
    \colhead{Ne3O2} \vline&
    \colhead{Median $z$} &
    \colhead{Median Mass} &
    \colhead{N} &
    \colhead{O32}}
    \startdata
    0.6 & 1.0 & 0.734 & 8.39 & 54 & -0.95$^{+0.09}_{-0.11}$ & 0.721 & 8.34 & 47 & 0.16$^{+0.03}_{-0.03}$ \\
    0.6 & 1.0 & 0.736 & 8.76 & 195 & -1.10$^{+0.04}_{-0.05}$ & 0.716 & 8.76 & 157 & 0.004$^{+0.03}_{-0.04}$ \\
    0.6 & 1.0 & 0.778 & 9.32 & 117 & -1.20$^{+0.04}_{-0.04}$ & 0.745 & 9.28 & 78 & -0.14$^{+0.03}_{-0.03}$ \\
    0.6 & 1.0 & 0.858 & 9.78 & 178 & -1.32$^{+0.04}_{-0.05}$ & 0.884 & 9.76 & 84 & -0.31$^{+0.06}_{-0.07}$ \\
    0.6 & 1.0 & 0.853 & 10.25 & 169 & -1.28$^{+0.08}_{-0.10}$ & 0.745 & 10.25 & 82 & -0.63$^{+0.13}_{-0.19}$ \\
    0.6 & 1.0 & 0.861 & 10.71 & 120 & -1.11$^{+0.10}_{-0.13}$ & 0.870 & 10.73 & 57 & -0.85$^{+0.18}_{-0.31}$
    \enddata
\end{deluxetable*}

\section{\neiii\ Relations} \label{sec:results}

\subsection{Ne3O2 and Stellar Mass}

Our primary objective in measuring the \neiii\ emission line is to compare its strength with the \oii\ line. The $log_{10}$(\neiii/\oii) ratio, hereafter Ne3O2, is a common proxy for gas-phase ionization and metallicity in galaxies. These two characteristics of galaxies have been shown to evolve from higher ionization and lower metallicity at the $z\sim2$ peak of cosmic star formation \citep{madau14} to lower ionization and higher metallicity in the local universe \citep[e.g.][]{steidel2014,strom2017}.

Furthermore, diagnostics such as the Mass-Excitation diagram \citep{juneau2014} demonstrate notable anticorrelations between the strengths of common excitation ratios, such as \oiii/H$\beta$, and increasing stellar mass among star-forming galaxies. Considering known SF-galaxy relations between stellar mass and gas-phase metallicity \citep[the Mass-Metallicity Relation, e.g.,][]{tremonti2004} and stellar mass and star formation rate \citep[e.g.,][]{noeske07} for a range of redshifts, it is useful to examine the Ne3O2 ratio as a function of stellar mass, both to explore the relation between ionization and mass and as a means for testing redshift evolution in the excitation ratio at fixed stellar mass.

\begin{figure}
    \centering
    \includegraphics[width=0.5\textwidth]{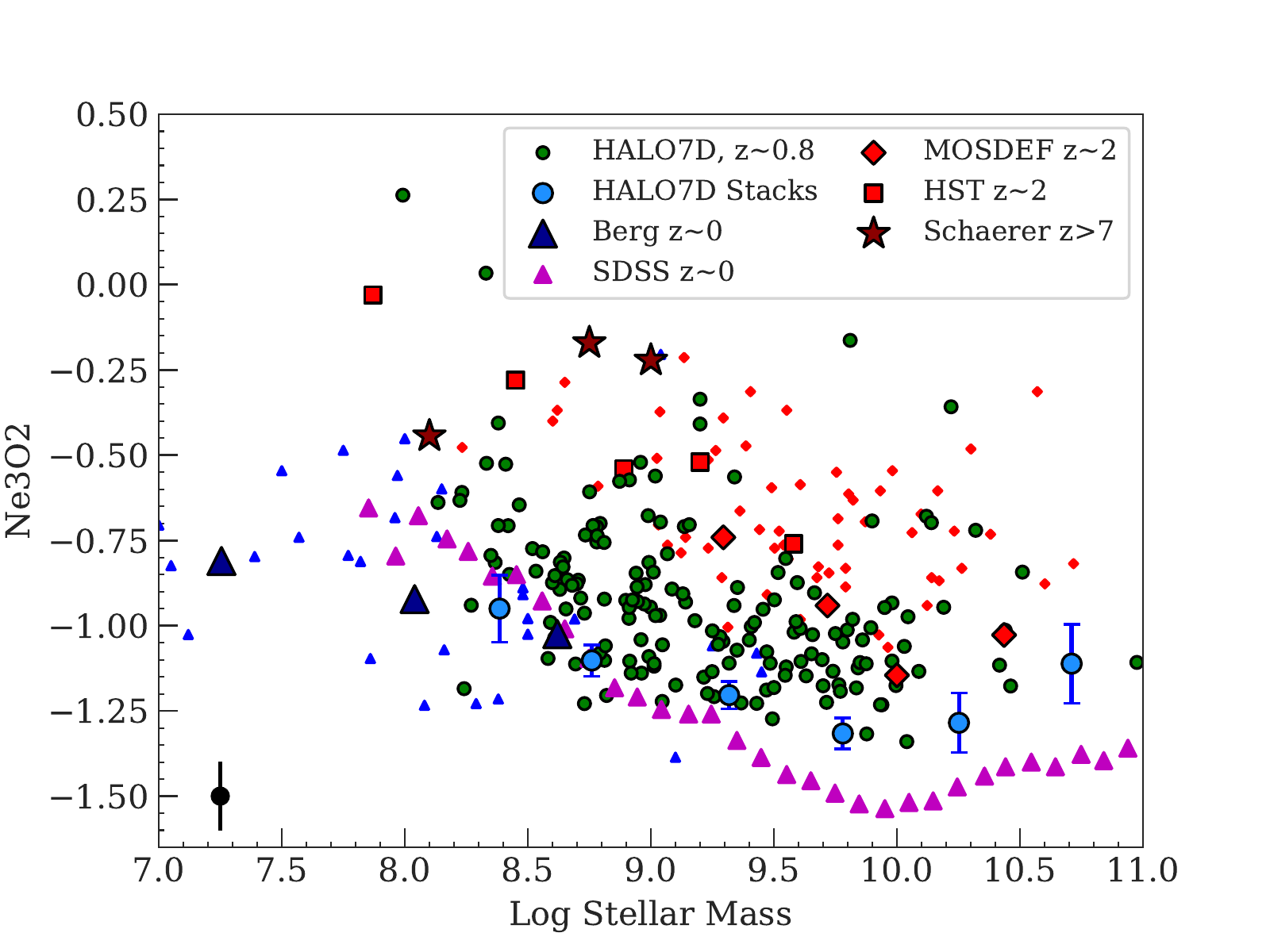}
    \caption{Ne3O2 vs. stellar mass for individual HALO7D galaxies (green circles) and composite HALO7D spectra (blue circles) compared with $z\sim2$ individual and composite measurements from J20 (black, red diamonds), local SDSS composite measurements from AM13 (purple triangles), and local low-mass individual detections (light blue triangles) and composite measures (dark triangles) from \citet{berg2012}. Measurements of three $z>7$ galaxies are shown as crimson stars \citep{schaerer2022}. Composite HALO7D spectra have lower Ne3O2 compared to the median individual detection at fixed stellar mass, suggesting the individual detections have higher excitation than the average for their mass and redshift. This relation holds in the $z\sim2$ and $z\sim0$ samples as well. Composite HALO7D spectra have similar Ne3O2 to $z\sim0$ stacks at low-mass, indicating that the Ne3O2 behavior doesn't evolve substantially from $z\sim1$ to $z\sim0$. The median error of individual HALO7D galaxies is given in black in the lower left.}
    \label{fig:ne3_m}
\end{figure}

Individual Ne3O2 measurements from HALO7D are shown as a function of stellar mass as green circles in Figure \ref{fig:ne3_m}, with masses from CANDELS catalogs (see \citet{pharo2022} for further description). Composite HALO7D spectra are shown with blue circles. The composite spectra bins sit at lower Ne3O2 than almost all individual detections with comparable stellar mass. This reflects the difference between the completeness limitations of the individual measurements, which are necessarily restricted to detection of the strongest emitters of this typically-faint emission line. The composite spectra, including information from a broader set of the line emitting population, suggest the average or ``typical'' behavior of line emitters at this redshift and stellar mass. A similar relationship between composite and individual measurements can be seen in the J20 and AM13 samples.

\subsection{Ne3O2 vs. O32}

For a subset of the \neon emitters, the \oiii\ emission lines are also detected. This subsample has a restricted redshift range, as the [O\textsc{iii}] lines move off the red end of the DEIMOS detection beyond $z=0.85$, and comparison of the Ne3O2 lines with [O\textsc{iii}] requires consideration of the dust extinction. However, the $log_{10}$(\oiii/\oii) ratio (hereafter O32) is another commonly used diagnostic for ionization and metallicity in galaxies, and so may add insight to Ne3O2 properties of the HALO7D sample. For the Ne3O2 galaxies with an [O\textsc{iii}] detection with S/N$>3$, we correct for dust extinction by measuring the Balmer decrement from the H$\beta$ and H$\gamma$ emission lines, also subject to the $S/N > 3$ criterion. For this correction, we used a \citet{cardelli1989} extinction law for nebular dust extinction calculations. The O32 sample properties are described in the right panel of Table \ref{tab:stack}.

\begin{figure*}
    \centering
    \begin{tabular}{cc}
    \includegraphics[width=0.5\textwidth]{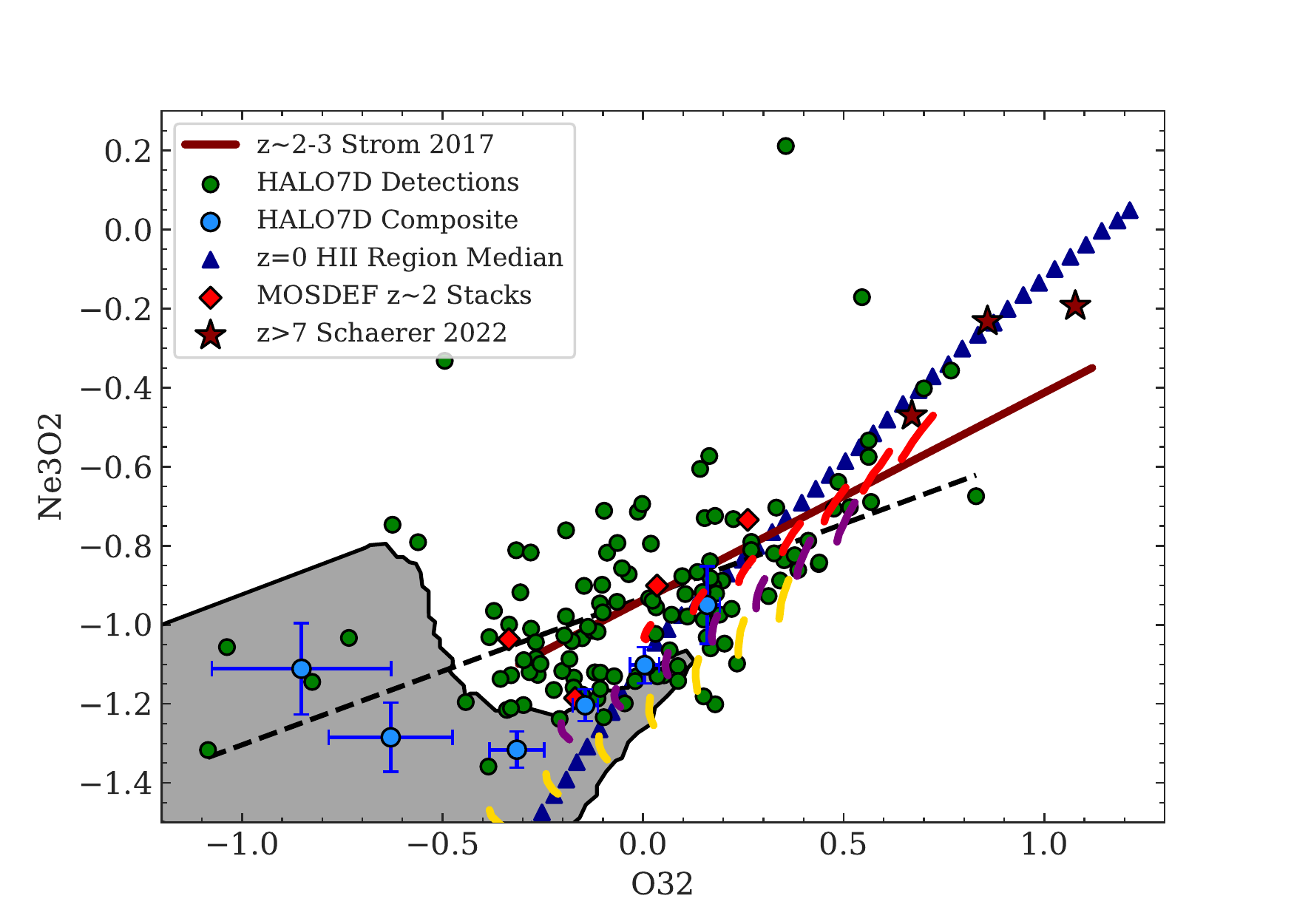} & \includegraphics[width=0.5\textwidth]{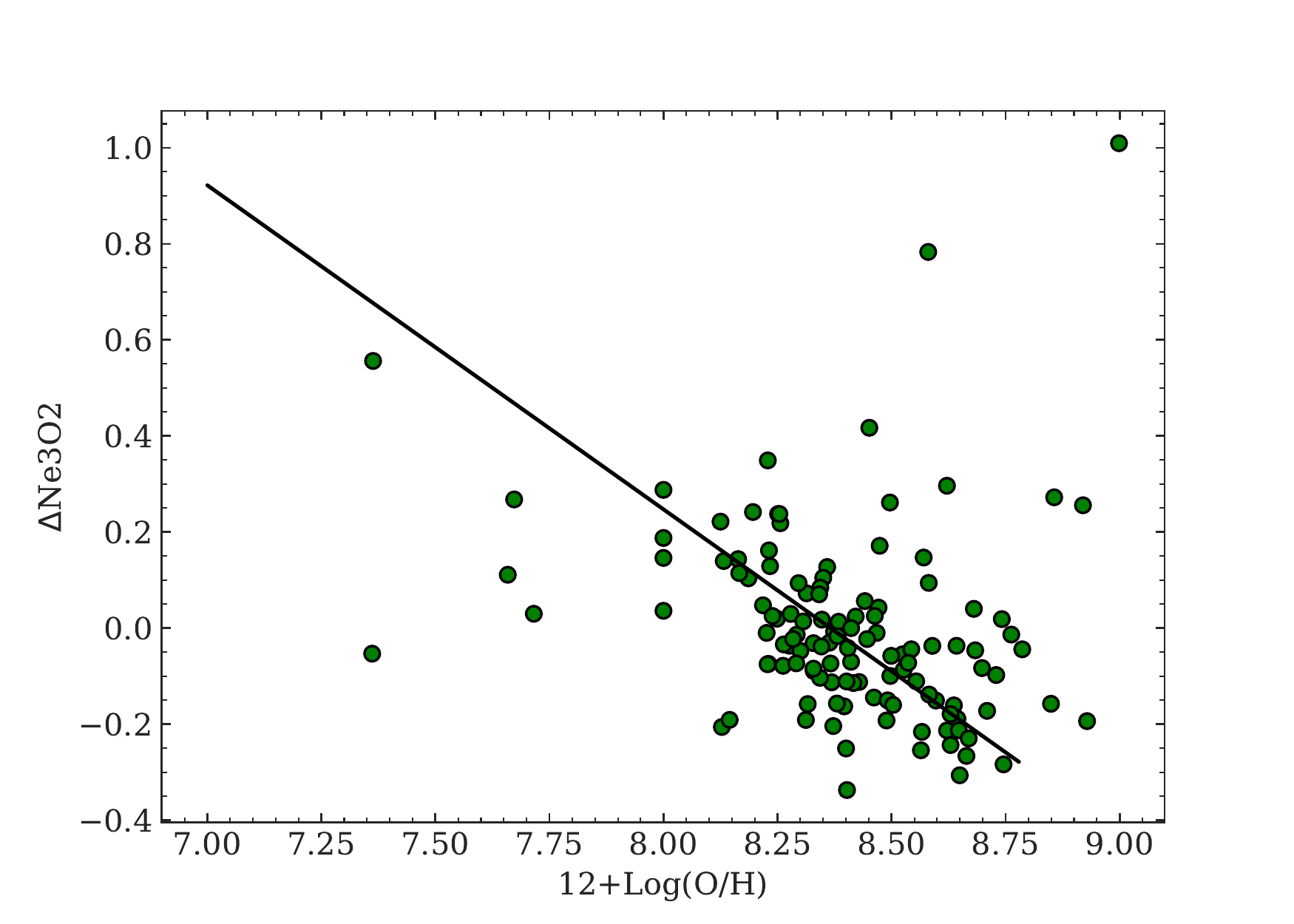}
    \end{tabular}
    \caption{\textit{Left:} Ne3O2 vs. O32 for individual HALO7D galaxies (green circles) and composite HALO7D spectra (blue circles) compared with $2<z<3$ composite spectra from J20 (red diamonds) and individual measurements from \citet{strom2017} (maroon line). We also include median values for local HII regions (blue triangles) assembled in J20 from \citet{pilyugin2016,croxall2016,torribio2016}. The gray shaded region indicates the 90\% boundary for local SDSS galaxies measured in \citet{strom2017}. The dashed black line shows a linear fit of the HALO7D points. Individual \neiii\ detections in HALO7D exhibit comparable ionization properties to (typically more massive) $2<z<3$ galaxies, but with high scatter that may be related to metallicity variations. The red, purple, and yellow curves give the BPASS models from \citet{strom2017}. \textit{Right:} The Ne3O2 offset from the linear Ne3O2-O32 fit as a function of the R23 metallicity.}
    \label{fig:ne3_o32}
\end{figure*}

Figure \ref{fig:ne3_o32} gives the Ne3O2 distribution as a function of the O32 ratio, with individual HALO7D detections shown in green and composite measurements shown in blue. As before, the individual measurements exhibit stronger emission characteristics concomitant with higher levels of ionization in the nebular gas compared with the composite measurements. In the following section, we analyze the Ne3O2 distributions and discuss their implications for the redshift evolution of ISM conditions in star-forming galaxies.

\section{Discussion and Conclusions} \label{sec:disc}

\subsection{Redshift Evolution from $z\sim2$}

In Figure \ref{fig:ne3_m}, the composite HALO7D spectra at $z\sim0.7-0.9$ closely track the low-$z$ composite spectra from AM13 at low mass, while at high mass, the Ne3O2 measurements sit between those measured at $z=0$ and $z\sim2$. This could suggest that the ionization and/or metallicity characteristics of low-mass star-forming galaxies do not evolve substantially from $z\sim1$ to $z=0$. However, at $\sm < 9$, the AM13 composites are biased toward starburst-SF galaxies, potentially biasing them to higher levels of Ne3O2 as well \citep{kashino2019}. To check for this bias, we also compare to the $z\sim0$ LVL sample, which was selected to probe representative main-sequence galaxies at low mass. We create stacked measurements from this sample in bins of stellar mass. The dark blue triangles in Figure \ref{fig:ne3_m} show that this sample closely matches the two HALO7D composites with $\sm < 9$. This comparison indicates a lack of ionization evolution from $z\sim1$ to $z=0$ more convincingly. It is also indicative of the completeness achieved by the deep HALO7D observations, which can probe low line emission comparable to $z=0$ observations out to a median redshift of $z=0.73$ for dwarf galaxies. It is more difficult to study the dwarf galaxy evolution out to $z\sim2-3$, where the faintness of low-mass line emitters makes their detection even more challenging. However, the lowest-mass MOSDEF bin, covering galaxies with masses $8.23 < \sm < 9.51$, nonetheless sits between 0.2 and 0.5 dex higher in Ne3O2 than any of the comparable mass bins from HALO7D. Individual detections in the two surveys show similar offsets. These results together indicate that the observed systematic decline in ionization parameter from $z\sim2$ to $z=0$ had largely ceased by $z\sim1$.

For mass bins $\sm > 9.5$, the HALO7D composites have higher Ne3O2 compared with the massive AM13 stacks by 0.1-0.2 dex. At stellar masses $\sm > 9$, where MOSDEF has comparable stacks from $z\sim2$, we see offsets between HALO7D and MOSDEF ranging from 0.4 to 0.1 dex. This indicates an ongoing decline in ionization in the higher-mass galaxies that could be driven by ongoing metallicity enrichment \citep{perezmontero2014} or declining rates of star formation in galaxies that have already built up stellar mass. 

To further explore ionization characteristics, we study the Ne3O2 vs O32 distribution in Figure \ref{fig:ne3_o32}. \citet{strom2017} used photoionization models to predict Ne3O2 and O32 ratios from BPASS stellar spectra \citep{stanway2016} for ranges of ionization parameter, gas-phase metallicity, and stellar metallicity. They show that the O32 ratio increases as a function of the ionization parameter, as well as with a reduction in stellar metallicity. O32 is less susceptible to changes in nebular metallicity except at high O32, but variation in the gas-phase metallicity by a factor from $Z_{neb} = 0.3Z_{\odot}$ to $Z_{neb} = 0.7Z_{\odot}$ can amount to a 0.2 dex drop in Ne3O2. Changing stellar population model metallicities from $Z_{\star} = 0.07Z_{\odot}$ (red track in Figure \ref{fig:ne3_o32}) to $Z_{\star}=0.56Z_{\odot}$ (purple track) can account for even larger drops in Ne3O2, from 0.2 to 0.5 dex. The analysis in J20 find similar trends with metallicity and ionization parameter, with variation in stellar metallicity able to account for vertical (Ne3O2) scatter in the distribution.

These models give a context to place the $z\sim0.7-0.8$ HALO7D galaxies. The broad trend of Ne3O2 increasing with higher O32 is a function of both ratios' sensitivity to increasing ionization parameter, and at given O32 values, the individual Ne3O2 measurements actually track the median from KBSS-MOSFIRE quite well. Similar Ne3O2 at given O32 among the two samples implies similar metallicity characteristics as well, and the median individual detection is thus also best described by the model with $Z_{\star} = 0.07Z_{\odot}$ at a range of ionization parameters. The scatter in individual detections is large, but can potentially be explained by variation in nebular metallicity. To check this, we do a simple linear fit to the Ne3O2-O32 distribution, then take the difference between each Ne3O2 value and the Ne3O2 predicted by the O32 value and the fit. This difference value correlates significantly with R23-derived metallicities\footnote{R23 = (\oiii +\oii)$/H\beta$, measured using the calibration from \citet{maiolino2008} with O\textsc{iii}5007/\oii$>3$ to break branch degeneracy, as in \citet{guo16a}}, with a Spearman correlation coefficient of -0.23 and $p=0.009$ (see right panel of Figure \ref{fig:ne3_o32}). This correlation suggests that lower metallicity leads to higher Ne3O2 scatter relative to the relation, and increased metallicity leads to lower Ne3O2. \citet{guo16a} observed increased metallicity scatter at given stellar mass among low-mass galaxies at similar redshift compared to the MZR of higher-mass galaxies, measured via strong-line methods. This potentially links Ne3O2 scatter to the sources of metallicity scatter in dwarf galaxies, which we will explore in a subsequent HALO7D paper.

Most of the composite spectra from HALO7D (blue circles) reside in the same region as low-$z$ SDSS galaxies (gray shaded region), except for the lowest-mass stacks, which are found outside the shaded 90\% region. Thus, the $\sm < 9$ stacks have higher O32 than typical local galaxies, but with Ne3O2 values that are still below the $z>2$ median. These non-local ratios can be explained by higher-than-local ionization parameters but with somewhat higher stellar metallicities. The two low-mass composite measurements correspond more closely to the $Z_{\star}=0.56Z_{\odot}$ (purple) track from \citet{strom2017}, suggesting, as in the Ne3O2-$M_{\star}$ plot, that the typical $z\sim0.7$ star-forming galaxy has metallicity characteristics already enriched to local values. The individual detections, however, provide a subset of galaxies whose emission characteristics are similar to $z>2$ galaxies whose ionizing conditions have not yet evolved, making this group a potentially interesting target for future analysis, since we are able to probe at lower redshift a population with both ISM and stellar mass characteristics comparable to high-$z$ galaxies.

\subsection{Comparison to $z>7$ Galaxies}

In Figures \ref{fig:ne3_m} and \ref{fig:ne3_o32}, we also include three $z>7$ galaxies identified in JWST Early Release Observations of the lensing cluster SMACS J0723.3-7327. \citet{schaerer2022} measured several emission line ratios from NIRSpec observations of the galaxies, including Ne3O2. The three have high Ne3O2 and O32 values and low stellar masses ($\sm \leq9$), and are marked with crimson stars in the figures. The emission lines ratios for the $z>7$ galaxies are on the extreme end of both the HALO7D and $z\sim2$ samples, but a handful of HALO7D galaxies have similar masses and emission characteristics. 

In addition to the emission ratios, the high-Ne3O2 dwarfs tend to share some other properties in common with the high-$z$ galaxies. They have high [O\textsc{iii}] equivalent width (EW$>100\AA$), compact size ($r \leq 2$ kpc), low gas-phase metallicity ($7 < 12+Log(O/H) < 8$), and high star formation rate surface density ($\Sigma_{SFR} > 0.1 M_{\odot}$ yr$^{-1}$ kpc$^{-2}$) compared to other HALO7D strong line emitters. These are all commonly thought to be characteristics of high-$z$ ionizing galaxies, and have been measured for these three galaxies in early analysis \citep{schaerer2022, rhoads2022}. Use of extreme low-$z$ galaxies as analogues of high-redshift UV-bright galaxies has been common practice \citep[e.g.][]{yang17}, and the properties listed above have been shown to be commonly found in low-redshift Lyman continuum leakers as well \citep{flury2022}. Compared to the overall distribution of HALO7D line emitters, the sample of individual Ne3O2 measurements is skewed significantly toward enhanced $\Sigma_{SFR}$, with almost all \neiii\ detections in galaxies with $\Sigma_{SFR} > 0.1 M_{\odot}$ yr$^{-1}$ kpc$^{-2}$, a major criterion used in \citet{flury2022} for Lyman continuum candidates.

This places the dwarf \neiii\ sample from HALO7D at a useful point in between high- and low-$z$ ionizing galaxy candidates. Ne3O2 has already seen use in constraining ionization in a $z\sim4.9$ gravitationally lensed massive galaxy \citep{witstok2021}, and as it becomes possible to obtain high-resolution spectroscopy of potentially-ionizing high-$z$ galaxies with JWST, we will be able to make direct comparisons of ISM conditions between dwarf galaxies at $z\sim0$, $z\sim1$, $z\sim2$, and $z>5$. Thus, the HALO7D sample will provide both an increased number of lower-redshift low-mass galaxies as potential analogues to high-$z$ sources, and by filling in the gap of low-mass ionizing galaxies at $z\sim1$, will continue to constrain the evolution in ionizing ISM properties in dwarf galaxies with redshift.

\begin{acknowledgments}
This research project was supported by NASA/ADAP grant number 80NSSC20K0443. D. Koo would like to acknowledge support from NSF grant AST-1615730. This research made use of Astropy, a community-developed core Python package for Astronomy (Astropy Collaboration et al. 2013, 2018). We recognize and acknowledge the significant cultural role and reverence that the summit of Maunakea has always had within the indigenous Hawaiian community. We are most fortunate to have the opportunity to use observations conducted from this mountain.
\end{acknowledgments}

%

\vspace{5mm}
\facilities{Keck(DEIMOS)}


\software{astropy \citep{astropy2022}}





\bibliography{refs}{}

\begin{thebibliography}{}
\expandafter\ifx\csname natexlab\endcsname\relax\def\natexlab#1{#1}\fi
\providecommand{\url}[1]{\href{#1}{#1}}
\providecommand{\dodoi}[1]{doi:~\href{http://doi.org/#1}{\nolinkurl{#1}}}
\providecommand{\doeprint}[1]{\href{http://ascl.net/#1}{\nolinkurl{http://ascl.net/#1}}}
\providecommand{\doarXiv}[1]{\href{https://arxiv.org/abs/#1}{\nolinkurl{https://arxiv.org/abs/#1}}}

\bibitem[{{Andrews} \& {Martini}(2013)}]{am2013}
{Andrews}, B.~H., \& {Martini}, P. 2013, \apj, 765, 140,
  \dodoi{10.1088/0004-637X/765/2/140}

\bibitem[{{Astropy Collaboration} {et~al.}(2022){Astropy Collaboration},
  {Price-Whelan}, {Lim}, {Earl}, {Starkman}, {Bradley}, {Shupe}, {Patil},
  {Corrales}, {Brasseur}, {N{\"o}the}, {Donath}, {Tollerud}, {Morris},
  {Ginsburg}, {Vaher}, {Weaver}, {Tocknell}, {Jamieson}, {van Kerkwijk},
  {Robitaille}, {Merry}, {Bachetti}, {G{\"u}nther}, {Aldcroft},
  {Alvarado-Montes}, {Archibald}, {B{\'o}di}, {Bapat}, {Barentsen},
  {Baz{\'a}n}, {Biswas}, {Boquien}, {Burke}, {Cara}, {Cara}, {Conroy},
  {Conseil}, {Craig}, {Cross}, {Cruz}, {D'Eugenio}, {Dencheva}, {Devillepoix},
  {Dietrich}, {Eigenbrot}, {Erben}, {Ferreira}, {Foreman-Mackey}, {Fox},
  {Freij}, {Garg}, {Geda}, {Glattly}, {Gondhalekar}, {Gordon}, {Grant},
  {Greenfield}, {Groener}, {Guest}, {Gurovich}, {Handberg}, {Hart},
  {Hatfield-Dodds}, {Homeier}, {Hosseinzadeh}, {Jenness}, {Jones}, {Joseph},
  {Kalmbach}, {Karamehmetoglu}, {Ka{\l}uszy{\'n}ski}, {Kelley}, {Kern},
  {Kerzendorf}, {Koch}, {Kulumani}, {Lee}, {Ly}, {Ma}, {MacBride}, {Maljaars},
  {Muna}, {Murphy}, {Norman}, {O'Steen}, {Oman}, {Pacifici}, {Pascual},
  {Pascual-Granado}, {Patil}, {Perren}, {Pickering}, {Rastogi}, {Roulston},
  {Ryan}, {Rykoff}, {Sabater}, {Sakurikar}, {Salgado}, {Sanghi}, {Saunders},
  {Savchenko}, {Schwardt}, {Seifert-Eckert}, {Shih}, {Jain}, {Shukla}, {Sick},
  {Simpson}, {Singanamalla}, {Singer}, {Singhal}, {Sinha}, {Sip{\H{o}}cz},
  {Spitler}, {Stansby}, {Streicher}, {{\v{S}}umak}, {Swinbank}, {Taranu},
  {Tewary}, {Tremblay}, {Val-Borro}, {Van Kooten}, {Vasovi{\'c}}, {Verma}, {de
  Miranda Cardoso}, {Williams}, {Wilson}, {Winkel}, {Wood-Vasey}, {Xue},
  {Yoachim}, {Zhang}, {Zonca}, \& {Astropy Project Contributors}}]{astropy2022}
{Astropy Collaboration}, {Price-Whelan}, A.~M., {Lim}, P.~L., {et~al.} 2022,
  \apj, 935, 167, \dodoi{10.3847/1538-4357/ac7c74}

\bibitem[{{Berg} {et~al.}(2012){Berg}, {Skillman}, {Marble}, {van Zee},
  {Engelbracht}, {Lee}, {Kennicutt}, {Calzetti}, {Dale}, \&
  {Johnson}}]{berg2012}
{Berg}, D.~A., {Skillman}, E.~D., {Marble}, A.~R., {et~al.} 2012, \apj, 754,
  98, \dodoi{10.1088/0004-637X/754/2/98}

\bibitem[{{Cardelli} {et~al.}(1989){Cardelli}, {Clayton}, \&
  {Mathis}}]{cardelli1989}
{Cardelli}, J.~A., {Clayton}, G.~C., \& {Mathis}, J.~S. 1989, \apj, 345, 245,
  \dodoi{10.1086/167900}

\bibitem[{{Cooper} {et~al.}(2012){Cooper}, {Newman}, {Davis}, {Finkbeiner}, \&
  {Gerke}}]{cooperspec2d2012}
{Cooper}, M.~C., {Newman}, J.~A., {Davis}, M., {Finkbeiner}, D.~P., \& {Gerke},
  B.~F. 2012, {spec2d: DEEP2 DEIMOS Spectral Pipeline}, Astrophysics Source
  Code Library, record ascl:1203.003.
\newblock \doeprint{1203.003}

\bibitem[{{Croxall} {et~al.}(2016){Croxall}, {Pogge}, {Berg}, {Skillman}, \&
  {Moustakas}}]{croxall2016}
{Croxall}, K.~V., {Pogge}, R.~W., {Berg}, D.~A., {Skillman}, E.~D., \&
  {Moustakas}, J. 2016, \apj, 830, 4, \dodoi{10.3847/0004-637X/830/1/4}

\bibitem[{{Cunningham} {et~al.}(2019{\natexlab{a}}){Cunningham}, {Deason},
  {Rockosi}, {Guhathakurta}, {Jennings}, {Kirby}, {Toloba}, \&
  {Barro}}]{cunningham2019a}
{Cunningham}, E.~C., {Deason}, A.~J., {Rockosi}, C.~M., {et~al.}
  2019{\natexlab{a}}, \apj, 876, 124, \dodoi{10.3847/1538-4357/ab16cb}

\bibitem[{{Cunningham} {et~al.}(2019{\natexlab{b}}){Cunningham}, {Deason},
  {Sanderson}, {Sohn}, {Anderson}, {Guhathakurta}, {Rockosi}, {van der Marel},
  {Loebman}, \& {Wetzel}}]{cunningham2019b}
{Cunningham}, E.~C., {Deason}, A.~J., {Sanderson}, R.~E., {et~al.}
  2019{\natexlab{b}}, \apj, 879, 120, \dodoi{10.3847/1538-4357/ab24cd}

\bibitem[{{Faber} {et~al.}(2003){Faber}, {Phillips}, {Kibrick}, {Alcott},
  {Allen}, {Burrous}, {Cantrall}, {Clarke}, {Coil}, {Cowley}, {Davis}, {Deich},
  {Dietsch}, {Gilmore}, {Harper}, {Hilyard}, {Lewis}, {McVeigh}, {Newman},
  {Osborne}, {Schiavon}, {Stover}, {Tucker}, {Wallace}, {Wei}, {Wirth}, \&
  {Wright}}]{faber03}
{Faber}, S.~M., {Phillips}, A.~C., {Kibrick}, R.~I., {et~al.} 2003, in Society
  of Photo-Optical Instrumentation Engineers (SPIE) Conference Series, Vol.
  4841, Instrument Design and Performance for Optical/Infrared Ground-based
  Telescopes, ed. M.~{Iye} \& A.~F.~M. {Moorwood}, 1657--1669,
  \dodoi{10.1117/12.460346}

\bibitem[{{Flury} {et~al.}(2022){Flury}, {Jaskot}, {Ferguson}, {Worseck},
  {Makan}, {Chisholm}, {Saldana-Lopez}, {Schaerer}, {McCandliss}, {Wang},
  {Ford}, {Heckman}, {Ji}, {Giavalisco}, {Amorin}, {Atek}, {Blaizot},
  {Borthakur}, {Carr}, {Castellano}, {Cristiani}, {De Barros}, {Dickinson},
  {Finkelstein}, {Fleming}, {Fontanot}, {Garel}, {Grazian}, {Hayes}, {Henry},
  {Mauerhofer}, {Micheva}, {Oey}, {Ostlin}, {Papovich}, {Pentericci},
  {Ravindranath}, {Rosdahl}, {Rutkowski}, {Santini}, {Scarlata}, {Teplitz},
  {Thuan}, {Trebitsch}, {Vanzella}, {Verhamme}, \& {Xu}}]{flury2022}
{Flury}, S.~R., {Jaskot}, A.~E., {Ferguson}, H.~C., {et~al.} 2022, \apjs, 260,
  1, \dodoi{10.3847/1538-4365/ac5331}

\bibitem[{{Grogin} {et~al.}(2011){Grogin}, {Kocevski}, {Faber}, {Ferguson},
  {Koekemoer}, {Riess}, {Acquaviva}, {Alexander}, {Almaini}, {Ashby}, {Barden},
  {Bell}, {Bournaud}, {Brown}, {Caputi}, {Casertano}, {Cassata}, {Castellano},
  {Challis}, {Chary}, {Cheung}, {Cirasuolo}, {Conselice}, {Roshan Cooray},
  {Croton}, {Daddi}, {Dahlen}, {Dav{\'e}}, {de Mello}, {Dekel}, {Dickinson},
  {Dolch}, {Donley}, {Dunlop}, {Dutton}, {Elbaz}, {Fazio}, {Filippenko},
  {Finkelstein}, {Fontana}, {Gardner}, {Garnavich}, {Gawiser}, {Giavalisco},
  {Grazian}, {Guo}, {Hathi}, {H{\"a}ussler}, {Hopkins}, {Huang}, {Huang},
  {Jha}, {Kartaltepe}, {Kirshner}, {Koo}, {Lai}, {Lee}, {Li}, {Lotz}, {Lucas},
  {Madau}, {McCarthy}, {McGrath}, {McIntosh}, {McLure}, {Mobasher},
  {Moustakas}, {Mozena}, {Nandra}, {Newman}, {Niemi}, {Noeske}, {Papovich},
  {Pentericci}, {Pope}, {Primack}, {Rajan}, {Ravindranath}, {Reddy}, {Renzini},
  {Rix}, {Robaina}, {Rodney}, {Rosario}, {Rosati}, {Salimbeni}, {Scarlata},
  {Siana}, {Simard}, {Smidt}, {Somerville}, {Spinrad}, {Straughn}, {Strolger},
  {Telford}, {Teplitz}, {Trump}, {van der Wel}, {Villforth}, {Wechsler},
  {Weiner}, {Wiklind}, {Wild}, {Wilson}, {Wuyts}, {Yan}, \& {Yun}}]{grogin11}
{Grogin}, N.~A., {Kocevski}, D.~D., {Faber}, S.~M., {et~al.} 2011, \apjs, 197,
  35, \dodoi{10.1088/0067-0049/197/2/35}

\bibitem[{{Guo} {et~al.}(2016){Guo}, {Koo}, {Lu}, {Forbes}, {Rafelski},
  {Trump}, {Amor{\'\i}n}, {Barro}, {Dav{\'e}}, {Faber}, {Hathi}, {Yesuf},
  {Cooper}, {Dekel}, {Guhathakurta}, {Kirby}, {Koekemoer},
  {P{\'e}rez-Gonz{\'a}lez}, {Lin}, {Newman}, {Primack}, {Rosario}, {Willmer},
  \& {Yan}}]{guo16a}
{Guo}, Y., {Koo}, D.~C., {Lu}, Y., {et~al.} 2016, \apj, 822, 103,
  \dodoi{10.3847/0004-637X/822/2/103}

\bibitem[{{Jeong} {et~al.}(2020){Jeong}, {Shapley}, {Sanders}, {Runco},
  {Topping}, {Reddy}, {Kriek}, {Coil}, {Mobasher}, {Siana}, {Shivaei},
  {Freeman}, {Azadi}, {Price}, {Leung}, {Fetherolf}, {de Groot}, {Zick},
  {Fornasini}, \& {Barro}}]{jeong2020}
{Jeong}, M.-S., {Shapley}, A.~E., {Sanders}, R.~L., {et~al.} 2020, \apjl, 902,
  L16, \dodoi{10.3847/2041-8213/abba7a}

\bibitem[{{Juneau} {et~al.}(2014){Juneau}, {Bournaud}, {Charlot}, {Daddi},
  {Elbaz}, {Trump}, {Brinchmann}, {Dickinson}, {Duc}, {Gobat}, {Jean-Baptiste},
  {Le Floc'h}, {Lehnert}, {Pacifici}, {Pannella}, \& {Schreiber}}]{juneau2014}
{Juneau}, S., {Bournaud}, F., {Charlot}, S., {et~al.} 2014, \apj, 788, 88,
  \dodoi{10.1088/0004-637X/788/1/88}

\bibitem[{{Kashino} \& {Inoue}(2019)}]{kashino2019}
{Kashino}, D., \& {Inoue}, A.~K. 2019, \mnras, 486, 1053,
  \dodoi{10.1093/mnras/stz881}

\bibitem[{{Kennicutt}(1998)}]{kennicutt98}
{Kennicutt}, Robert~C., J. 1998, \araa, 36, 189,
  \dodoi{10.1146/annurev.astro.36.1.189}

\bibitem[{{Kewley} {et~al.}(2001){Kewley}, {Dopita}, {Sutherland}, {Heisler},
  \& {Trevena}}]{kewley01}
{Kewley}, L.~J., {Dopita}, M.~A., {Sutherland}, R.~S., {Heisler}, C.~A., \&
  {Trevena}, J. 2001, \apj, 556, 121, \dodoi{10.1086/321545}

\bibitem[{{Kewley} {et~al.}(2019){Kewley}, {Nicholls}, \&
  {Sutherland}}]{kewley19}
{Kewley}, L.~J., {Nicholls}, D.~C., \& {Sutherland}, R.~S. 2019, \araa, 57,
  511, \dodoi{10.1146/annurev-astro-081817-051832}

\bibitem[{{Koekemoer} {et~al.}(2011){Koekemoer}, {Faber}, {Ferguson}, {Grogin},
  {Kocevski}, {Koo}, {Lai}, {Lotz}, {Lucas}, {McGrath}, {Ogaz}, {Rajan},
  {Riess}, {Rodney}, {Strolger}, {Casertano}, {Castellano}, {Dahlen},
  {Dickinson}, {Dolch}, {Fontana}, {Giavalisco}, {Grazian}, {Guo}, {Hathi},
  {Huang}, {van der Wel}, {Yan}, {Acquaviva}, {Alexander}, {Almaini}, {Ashby},
  {Barden}, {Bell}, {Bournaud}, {Brown}, {Caputi}, {Cassata}, {Challis},
  {Chary}, {Cheung}, {Cirasuolo}, {Conselice}, {Roshan Cooray}, {Croton},
  {Daddi}, {Dav{\'e}}, {de Mello}, {de Ravel}, {Dekel}, {Donley}, {Dunlop},
  {Dutton}, {Elbaz}, {Fazio}, {Filippenko}, {Finkelstein}, {Frazer}, {Gardner},
  {Garnavich}, {Gawiser}, {Gruetzbauch}, {Hartley}, {H{\"a}ussler},
  {Herrington}, {Hopkins}, {Huang}, {Jha}, {Johnson}, {Kartaltepe},
  {Khostovan}, {Kirshner}, {Lani}, {Lee}, {Li}, {Madau}, {McCarthy},
  {McIntosh}, {McLure}, {McPartland}, {Mobasher}, {Moreira}, {Mortlock},
  {Moustakas}, {Mozena}, {Nandra}, {Newman}, {Nielsen}, {Niemi}, {Noeske},
  {Papovich}, {Pentericci}, {Pope}, {Primack}, {Ravindranath}, {Reddy},
  {Renzini}, {Rix}, {Robaina}, {Rosario}, {Rosati}, {Salimbeni}, {Scarlata},
  {Siana}, {Simard}, {Smidt}, {Snyder}, {Somerville}, {Spinrad}, {Straughn},
  {Telford}, {Teplitz}, {Trump}, {Vargas}, {Villforth}, {Wagner}, {Wandro},
  {Wechsler}, {Weiner}, {Wiklind}, {Wild}, {Wilson}, {Wuyts}, \&
  {Yun}}]{koekemoer11}
{Koekemoer}, A.~M., {Faber}, S.~M., {Ferguson}, H.~C., {et~al.} 2011, \apjs,
  197, 36, \dodoi{10.1088/0067-0049/197/2/36}

\bibitem[{{Kriek} {et~al.}(2015){Kriek}, {Shapley}, {Reddy}, {Siana}, {Coil},
  {Mobasher}, {Freeman}, {de Groot}, {Price}, {Sanders}, {Shivaei}, {Brammer},
  {Momcheva}, {Skelton}, {van Dokkum}, {Whitaker}, {Aird}, {Azadi}, {Kassis},
  {Bullock}, {Conroy}, {Dav{\'e}}, {Kere{\v{s}}}, \& {Krumholz}}]{kriek2015}
{Kriek}, M., {Shapley}, A.~E., {Reddy}, N.~A., {et~al.} 2015, \apjs, 218, 15,
  \dodoi{10.1088/0067-0049/218/2/15}

\bibitem[{{Levesque} \& {Richardson}(2014)}]{levesque2014}
{Levesque}, E.~M., \& {Richardson}, M. L.~A. 2014, \apj, 780, 100,
  \dodoi{10.1088/0004-637X/780/1/100}

\bibitem[{{Madau} \& {Dickinson}(2014)}]{madau14}
{Madau}, P., \& {Dickinson}, M. 2014, \araa, 52, 415,
  \dodoi{10.1146/annurev-astro-081811-125615}

\bibitem[{{Maiolino} {et~al.}(2008){Maiolino}, {Nagao}, {Grazian}, {Cocchia},
  {Marconi}, {Mannucci}, {Cimatti}, {Pipino}, {Ballero}, {Calura}, {Chiappini},
  {Fontana}, {Granato}, {Matteucci}, {Pastorini}, {Pentericci}, {Risaliti},
  {Salvati}, \& {Silva}}]{maiolino2008}
{Maiolino}, R., {Nagao}, T., {Grazian}, A., {et~al.} 2008, \aap, 488, 463,
  \dodoi{10.1051/0004-6361:200809678}

\bibitem[{{Muzzin} {et~al.}(2013){Muzzin}, {Marchesini}, {Stefanon}, {Franx},
  {Milvang-Jensen}, {Dunlop}, {Fynbo}, {Brammer}, {Labb{\'e}}, \& {van
  Dokkum}}]{muzzin13}
{Muzzin}, A., {Marchesini}, D., {Stefanon}, M., {et~al.} 2013, \apjs, 206, 8,
  \dodoi{10.1088/0067-0049/206/1/8}

\bibitem[{{Nagao} {et~al.}(2006){Nagao}, {Maiolino}, \& {Marconi}}]{nagao2006}
{Nagao}, T., {Maiolino}, R., \& {Marconi}, A. 2006, \aap, 459, 85,
  \dodoi{10.1051/0004-6361:20065216}

\bibitem[{{Newman} {et~al.}(2013){Newman}, {Cooper}, {Davis}, {Faber}, {Coil},
  {Guhathakurta}, {Koo}, {Phillips}, {Conroy}, {Dutton}, {Finkbeiner}, {Gerke},
  {Rosario}, {Weiner}, {Willmer}, {Yan}, {Harker}, {Kassin}, {Konidaris},
  {Lai}, {Madgwick}, {Noeske}, {Wirth}, {Connolly}, {Kaiser}, {Kirby},
  {Lemaux}, {Lin}, {Lotz}, {Luppino}, {Marinoni}, {Matthews}, {Metevier}, \&
  {Schiavon}}]{newman13}
{Newman}, J.~A., {Cooper}, M.~C., {Davis}, M., {et~al.} 2013, \apjs, 208, 5,
  \dodoi{10.1088/0067-0049/208/1/5}

\bibitem[{{Noeske} {et~al.}(2007){Noeske}, {Weiner}, {Faber}, {Papovich},
  {Koo}, {Somerville}, {Bundy}, {Conselice}, {Newman}, {Schiminovich}, {Le
  Floc'h}, {Coil}, {Rieke}, {Lotz}, {Primack}, {Barmby}, {Cooper}, {Davis},
  {Ellis}, {Fazio}, {Guhathakurta}, {Huang}, {Kassin}, {Martin}, {Phillips},
  {Rich}, {Small}, {Willmer}, \& {Wilson}}]{noeske07}
{Noeske}, K.~G., {Weiner}, B.~J., {Faber}, S.~M., {et~al.} 2007, \apjl, 660,
  L43, \dodoi{10.1086/517926}

\bibitem[{{Oke} \& {Gunn}(1983)}]{oke83}
{Oke}, J.~B., \& {Gunn}, J.~E. 1983, \apj, 266, 713, \dodoi{10.1086/160817}

\bibitem[{P\'erez-Montero(2014)}]{perezmontero2014}
P\'erez-Montero, E. 2014, Monthly Notices of the Royal Astronomical Society,
  441, 2663, \dodoi{10.1093/mnras/stu753}

\bibitem[{{Pharo} {et~al.}(2022){Pharo}, {Guo}, {Calvo}, {Carleton}, {Faber},
  {Guhathakurta}, {Kassin}, {Koo}, {Lonergan}, {Teppala}, {Wang}, {Yesuf},
  {Bian}, {Dav{\'e}}, {Forbes}, {Keres}, {Perez-Gonzalez}, {Martin}, {Puleo},
  {Williams}, \& {Winningham}}]{pharo2022}
{Pharo}, J., {Guo}, Y., {Calvo}, G.~B., {et~al.} 2022, \apjs, 261, 12,
  \dodoi{10.3847/1538-4365/ac6cdf}

\bibitem[{{Pilyugin} \& {Grebel}(2016)}]{pilyugin2016}
{Pilyugin}, L.~S., \& {Grebel}, E.~K. 2016, \mnras, 457, 3678,
  \dodoi{10.1093/mnras/stw238}

\bibitem[{{Rhoads} {et~al.}(2022){Rhoads}, {Wold}, {Harish}, {Kim}, {Pharo},
  {Malhotra}, {Gabrielpillai}, {Jiang}, \& {Yang}}]{rhoads2022}
{Rhoads}, J.~E., {Wold}, I. G.~B., {Harish}, S., {et~al.} 2022, arXiv e-prints,
  arXiv:2207.13020.
\newblock \doarXiv{2207.13020}

\bibitem[{{Rudie} {et~al.}(2012){Rudie}, {Steidel}, {Trainor}, {Rakic},
  {Bogosavljevi{\'c}}, {Pettini}, {Reddy}, {Shapley}, {Erb}, \&
  {Law}}]{rudie2012}
{Rudie}, G.~C., {Steidel}, C.~C., {Trainor}, R.~F., {et~al.} 2012, \apj, 750,
  67, \dodoi{10.1088/0004-637X/750/1/67}

\bibitem[{{Sanders} {et~al.}(2016){Sanders}, {Shapley}, {Kriek}, {Reddy},
  {Freeman}, {Coil}, {Siana}, {Mobasher}, {Shivaei}, {Price}, \& {de
  Groot}}]{sanders2016}
{Sanders}, R.~L., {Shapley}, A.~E., {Kriek}, M., {et~al.} 2016, \apj, 816, 23,
  \dodoi{10.3847/0004-637X/816/1/23}

\bibitem[{{Sanders} {et~al.}(2018){Sanders}, {Shapley}, {Kriek}, {Freeman},
  {Reddy}, {Siana}, {Coil}, {Mobasher}, {Dav{\'e}}, {Shivaei}, {Azadi},
  {Price}, {Leung}, {Fetherolf}, {de Groot}, {Zick}, {Fornasini}, \&
  {Barro}}]{sanders2018}
---. 2018, \apj, 858, 99, \dodoi{10.3847/1538-4357/aabcbd}

\bibitem[{{Schaerer} {et~al.}(2022){Schaerer}, {Marques-Chaves}, {Barrufet},
  {Oesch}, {Izotov}, {Naidu}, {Guseva}, \& {Brammer}}]{schaerer2022}
{Schaerer}, D., {Marques-Chaves}, R., {Barrufet}, L., {et~al.} 2022, arXiv
  e-prints, arXiv:2207.10034.
\newblock \doarXiv{2207.10034}

\bibitem[{{Shapley} {et~al.}(2019){Shapley}, {Sanders}, {Shao}, {Reddy},
  {Kriek}, {Coil}, {Mobasher}, {Siana}, {Shivaei}, {Freeman}, {Azadi}, {Price},
  {Leung}, {Fetherolf}, {de Groot}, {Zick}, {Fornasini}, \&
  {Barro}}]{shapley2019}
{Shapley}, A.~E., {Sanders}, R.~L., {Shao}, P., {et~al.} 2019, \apjl, 881, L35,
  \dodoi{10.3847/2041-8213/ab385a}

\bibitem[{{Stanway} {et~al.}(2016){Stanway}, {Eldridge}, \&
  {Becker}}]{stanway2016}
{Stanway}, E.~R., {Eldridge}, J.~J., \& {Becker}, G.~D. 2016, \mnras, 456, 485,
  \dodoi{10.1093/mnras/stv2661}

\bibitem[{{Steidel} {et~al.}(2014){Steidel}, {Rudie}, {Strom}, {Pettini},
  {Reddy}, {Shapley}, {Trainor}, {Erb}, {Turner}, {Konidaris}, {Kulas}, {Mace},
  {Matthews}, \& {McLean}}]{steidel2014}
{Steidel}, C.~C., {Rudie}, G.~C., {Strom}, A.~L., {et~al.} 2014, \apj, 795,
  165, \dodoi{10.1088/0004-637X/795/2/165}

\bibitem[{{Strom} {et~al.}(2017){Strom}, {Steidel}, {Rudie}, {Trainor},
  {Pettini}, \& {Reddy}}]{strom2017}
{Strom}, A.~L., {Steidel}, C.~C., {Rudie}, G.~C., {et~al.} 2017, \apj, 836,
  164, \dodoi{10.3847/1538-4357/836/2/164}

\bibitem[{{Tacchella} {et~al.}(2020){Tacchella}, {Forbes}, \&
  {Caplar}}]{tacchella2020}
{Tacchella}, S., {Forbes}, J.~C., \& {Caplar}, N. 2020, \mnras, 497, 698,
  \dodoi{10.1093/mnras/staa1838}

\bibitem[{{Tacchella} {et~al.}(2022){Tacchella}, {Conroy}, {Faber}, {Johnson},
  {Leja}, {Barro}, {Cunningham}, {Deason}, {Guhathakurta}, {Guo}, {Hernquist},
  {Koo}, {McKinnon}, {Rockosi}, {Speagle}, {van Dokkum}, \&
  {Yesuf}}]{tacchella2022}
{Tacchella}, S., {Conroy}, C., {Faber}, S.~M., {et~al.} 2022, \apj, 926, 134,
  \dodoi{10.3847/1538-4357/ac449b}

\bibitem[{{Toribio San Cipriano} {et~al.}(2016){Toribio San Cipriano},
  {Garc{\'\i}a-Rojas}, {Esteban}, {Bresolin}, \& {Peimbert}}]{torribio2016}
{Toribio San Cipriano}, L., {Garc{\'\i}a-Rojas}, J., {Esteban}, C., {Bresolin},
  F., \& {Peimbert}, M. 2016, \mnras, 458, 1866, \dodoi{10.1093/mnras/stw397}

\bibitem[{{Tremonti} {et~al.}(2004){Tremonti}, {Heckman}, {Kauffmann},
  {Brinchmann}, {Charlot}, {White}, {Seibert}, {Peng}, {Schlegel}, {Uomoto},
  {Fukugita}, \& {Brinkmann}}]{tremonti2004}
{Tremonti}, C.~A., {Heckman}, T.~M., {Kauffmann}, G., {et~al.} 2004, \apj, 613,
  898, \dodoi{10.1086/423264}

\bibitem[{{Wang} {et~al.}(2022){Wang}, {Kassin}, {Faber}, {Koo}, {Cunningham},
  {Yesuf}, {Barro}, {Guhathakurta}, {Weiner}, {de la Vega}, {Guo}, {Heckman},
  {Pacifici}, {Wang}, \& {Welker}}]{wang2022}
{Wang}, W., {Kassin}, S.~A., {Faber}, S.~M., {et~al.} 2022, \apj, 930, 146,
  \dodoi{10.3847/1538-4357/ac6592}

\bibitem[{{Witstok} {et~al.}(2021){Witstok}, {Smit}, {Maiolino}, {Curti},
  {Laporte}, {Massey}, {Richard}, \& {Swinbank}}]{witstok2021}
{Witstok}, J., {Smit}, R., {Maiolino}, R., {et~al.} 2021, \mnras, 508, 1686,
  \dodoi{10.1093/mnras/stab2591}

\bibitem[{{Yang} {et~al.}(2017){Yang}, {Malhotra}, {Gronke}, {Rhoads},
  {Leitherer}, {Wofford}, {Jiang}, {Dijkstra}, {Tilvi}, \& {Wang}}]{yang17}
{Yang}, H., {Malhotra}, S., {Gronke}, M., {et~al.} 2017, \apj, 844, 171,
  \dodoi{10.3847/1538-4357/aa7d4d}

\bibitem[{{Yesuf} {et~al.}(2017){Yesuf}, {Koo}, {Faber}, {Prochaska}, {Guo},
  {Liu}, {Cunningham}, {Coil}, \& {Guhathakurta}}]{yesuf17}
{Yesuf}, H.~M., {Koo}, D.~C., {Faber}, S.~M., {et~al.} 2017, \apj, 841, 83,
  \dodoi{10.3847/1538-4357/aa6fae}

\bibitem[{{Zeimann} {et~al.}(2015){Zeimann}, {Ciardullo}, {Gebhardt},
  {Gronwall}, {Hagen}, {Trump}, {Bridge}, {Luo}, \& {Schneider}}]{zeimann2015}
{Zeimann}, G.~R., {Ciardullo}, R., {Gebhardt}, H., {et~al.} 2015, \apj, 798,
  29, \dodoi{10.1088/0004-637X/798/1/29}

\end{thebibliography}
\bibliographystyle{aasjournal}



\end{document}